\documentclass[]{aa}

\usepackage{graphicx}
\usepackage{txfonts}
\usepackage[dvipsnames]{xcolor}
\usepackage{hyperref}
\hypersetup{
    final=true,
    pageanchor=true,
    colorlinks=true,
    breaklinks=true,
    linkcolor=blue,
    citecolor=blue,
    urlcolor=blue,
    pdfpagemode=UseNone,
    pdftitle={},
    pdfauthor={E. G. Broock et al.},
    pdfsubject={Solar Physics},
    pdfkeywords={Sun:activity - Sun:helioseismology - Sun:oscillations - Sun:sunspots - Sun:UV radiation}}

\begin{document} 

\newcommand{\andres}[1]{\textcolor{ForestGreen}{#1}}
\newcommand{\tobias}[1]{\textcolor{blue}{#1}}
\newcommand{\corr}[1]{\textcolor{red}{#1}}
\newcommand{\elena}[1]{\textcolor{BurntOrange}{#1}}

   \title{Performance of solar far-side active regions neural detection}

   \subtitle{}

   \author{E. G. Broock\inst{\ref{inst1},\ref{inst2}}
          \and
          T. Felipe\inst{\ref{inst1},\ref{inst2}}    
          \and
          A. Asensio Ramos\inst{\ref{inst1},\ref{inst2}}
          }

   \institute{Instituto de Astrof\'{\i}sica de Canarias, 38205, C/ V\'{\i}a L{\'a}ctea, s/n, La Laguna, Tenerife, Spain\label{inst1}
         \and
             Departamento de Astrof\'{\i}sica, Universidad de La Laguna, 38205, La Laguna, Tenerife, Spain\label{inst2} 
             }

   \date{Received xxx, aaaa; accepted yyy, aaaa}

 
  \abstract
   {Far-side helioseismology is a technique used to infer the presence of active regions in the far hemisphere of the Sun based on the interpretation of oscillations measured in the near hemisphere. A neural network has been recently developed to improve the sensitivity of the seismic maps to the presence of far-side active regions.}
   {Our aim is to evaluate the performance of the new neural network approach and to thoroughly compare it with the standard method commonly applied to predict far-side active regions from seismic measurements.}
   {We have computed the predictions of active regions using the neural network and the standard approach from five years of far-side seismic maps as a function of the selected threshold in the signatures of the detections. The results have been compared with direct extreme ultraviolet observations of the far hemisphere acquired with the Solar Terrestrial Relations Observatory (STEREO).}
   {We have confirmed the improved sensitivity of the neural network to the presence of far-side active regions. Approximately 96\% of the active regions identified by the standard method with a strength above the threshold commonly employed by previous analyses are related to locations with enhanced extreme ultraviolet emission. For this threshold, the false positive ratio is 3.75\%. For an equivalent false positive ratio, the neural network produces 47\% more true detections.
   Weaker active regions can be detected by relaxing the threshold in their seismic signature. For almost all the range of thresholds, the performance of the neural network is superior to the standard approach, delivering a higher number of confirmed detections and a lower rate of false positives.}
   {The neural network is a promising approach to improve the interpretation of the seismic maps provided by local helioseismic techniques. Additionally, refined predictions of magnetic activity in the non-visible solar hemisphere can play a significant role in space weather forecasting.}


   \maketitle
%

\section{Introduction}\label{introduction}
Helioseismology is one of the few available techniques that can be used to infer the 
properties of the solar interior. It is based on the analysis of the oscillations on the surface. The first studies in 
this field were focused on the interpretation of the eigenfrequencies of the resonant modes. They led to remarkable results, such as insight on the solar interior rotation \citep{Christensen-Dalsgaard2002}. Despite the success, the study of the eigenfrequencies only allows the inference of the global properties of the Sun. To bypass this limitation, the
new field of "local helioseismology" has been developed over the last three decades \citep{Braun+etal1987, Hill1988, Braun+etal1992, Duvall+etal1993}. Local helioseismology focuses on the interpretation of not only eigenfrequencies, but the whole wave field, with the aim of exploring localized structures below the surface.

The interest on local helioseismology led to the development of a number of techniques
and applications \citep[see][for a review]{Gizon+Birch2005}. One of those is helioseismic holography \citep{Lindsey+Braun1990}, a technique based on the fundamental
idea that the wave field at the solar surface has the signature of the wave field of any region
in the solar interior at any given time. Phase-sensitivity holography is a particular case of local 
helioseismology used to infer time-travel perturbations. A detailed description of helioseismic holography, 
and of phase-sensitivity holography in particular, can be found in \citet{Lindsey+Braun2000b}.

Far-side imaging is an application of phase-sensitivity holography with the purpose
of detecting active regions in the non-visible hemisphere of the Sun \citep{Lindsey+Braun2000, Braun+Lindsey2001}. To this end, it uses data from a "pupil" 
(region of the solar surface where the wave field is observed) on the near-side of the Sun to infer the properties of a "focus point" on the far-side. Similar studies of far-side active regions have been performed using time-distance helioseismology instead of helioseismic holography \citep{Duvall+Kosovichev2001,Zhao2007, Ilonidis+etal2009}. The success of these methods relies on the transparency of 
the solar interior to seismic waves from a range of frequencies. These waves are reflected on the 
top layers of the solar interior and are refracted back to the surface far from the
original source leading to a "bounce-like" propagation. At regions with strong magnetic field concentrations, such as sunspots, the solar surface is depressed (an effect known as Wilson depression) and waves are reflected at a deeper layer \citep{Lindsey+etal2010, Schunker+etal2013, Felipe+etal2017b}. This shortening of the wave path can be detected in far-side seismic maps as a reduction in the travel time. Far-side imaging relies on multiple-skip waves, 
using waves that are reflected on the surface at least once on their propagation from the focus point 
to the pupil or vice versa. Recently, \cite{Zhao+etal2019} have developed new schemes to perform time-distance far-side analyses including waves whose paths follow a higher number of skips. 
\begin{figure*}[!tbp]
  \centering
  {\includegraphics[height=1.3in]{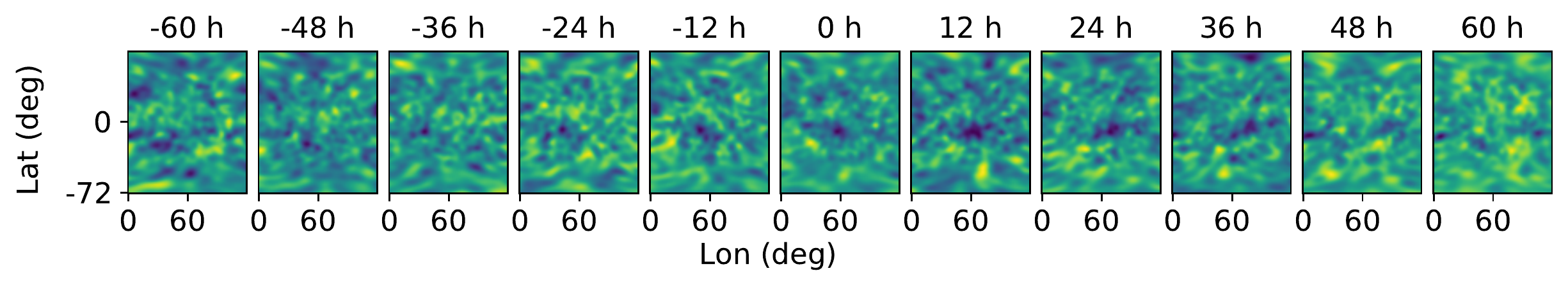}}
  {\includegraphics[height=2in]{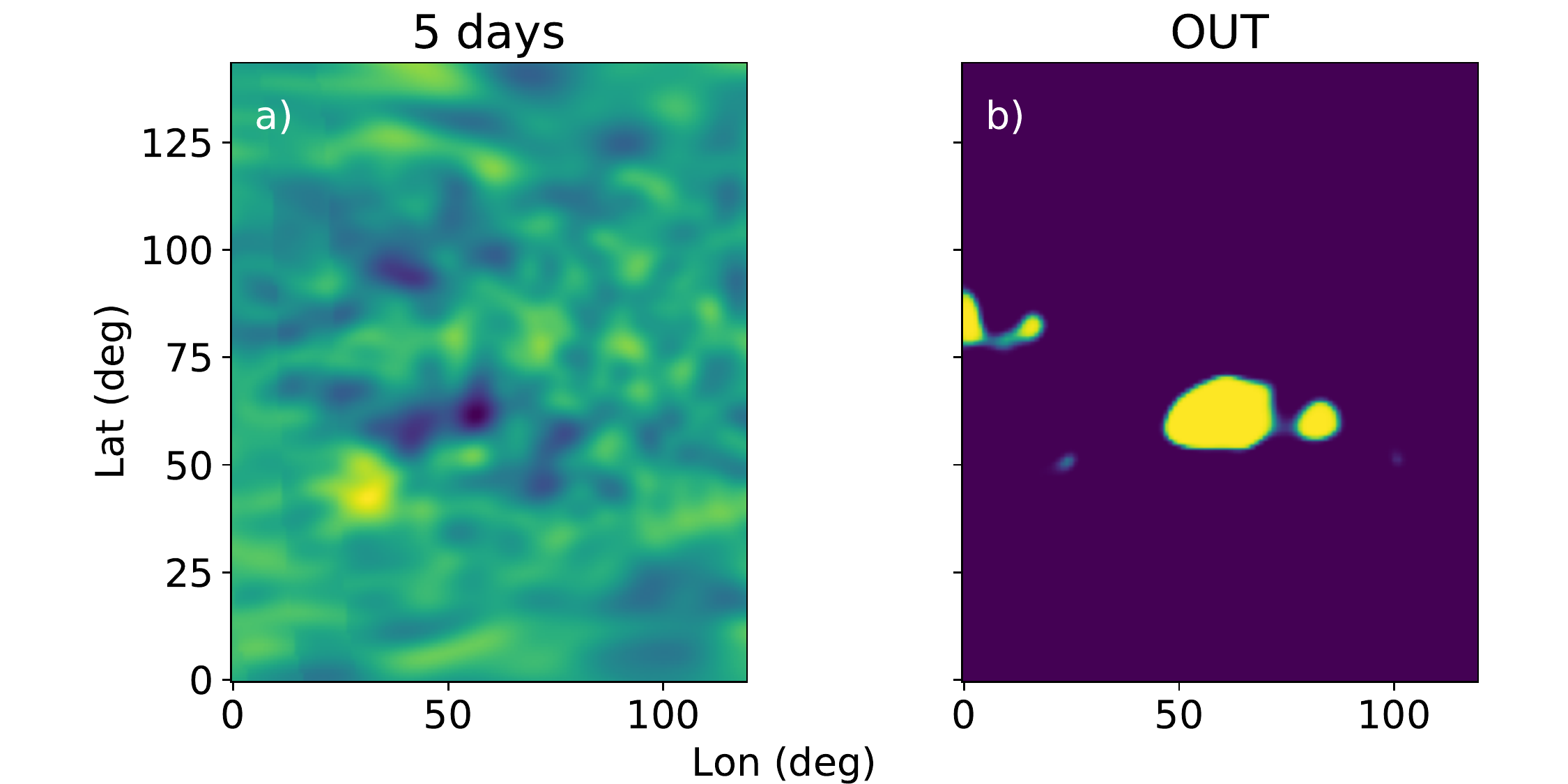}}
  \caption{Top: eleven phase-shift maps used as input to FarNet. Each map is computed with 24 h of Doppler observations. The time cadence of the seismic maps is 12 h. The colour scale indicates the phase shift, ranging from -0.28 rad (dark blue) to 0.17 rad (yellow). Bottom: five days cumulative phase-shift map (a) computed for
  the central time of the series (input time 0 h, December 25, 2013) and prediction of FarNet for the same time (b) using the eleven phase-shift maps.}
  \label{eleven}
\end{figure*}

The space weather on the Solar System and its influence on the biological and technological systems on Earth is a direct consequence of the activity on the solar surface. Learning about and preventing these effects is the main goal of space weather forecasting. Far-side helioseismology is a key ingredient for the forecasting because it
allows the detection of active regions that could represent a potential risk before they rotate into the visible solar hemisphere. Moreover, the inclusion of far-side active regions has been proven to greatly improve the performance of space weather forecast \citep{Fontenla+etal2009, Arge+etal2013}. See \cite{Lindsey+Braun2017} for a discussion of the applications of far-side seismic imaging in space weather.

Although promising and successful, all seismic techniques for the detection of far-side active regions are only able to detect the strongest active regions, leaving far-side fainter activity unprobed \citep{GonzalezHernandez+etal2007,Liewer+etal2014,Liewer+etal2017}. In an
effort to improve the detection capabilities, \citet{Felipe+Asensio2019} leveraged deep learning to develop a convolutional neural network (CNN) that improves the sensitivity of the method. We use FarNet to refer to this neural network hereafter.

The main goal of this paper is to analyse in depth FarNet, comparing its performance on the detection of active regions detection with standard far-side helioseismic techniques. We have employed direct Solar Terrestrial Relations Observatory \citep[STEREO,][]{Kaiser2004} far-side EUV images to test the reliability of the network predictions. The paper is organized as follows: Sect. \ref{detec} explains the data used to obtain the predictions of far-side active regions and gives insight about FarNet and the standard seismic method, Sect. \ref{STEREO} shows the data that we used as proxy of active regions for the comparison, Sect. \ref{comparison} describes the comparison between both methods, and Sect. \ref{disc} presents the discussion of the results.

\section{Detection of far-side active regions}\label{detec}

\subsection{Far-side seismic maps}\label{maps}
The fundamental data for the seismological identification of far-side active regions are phase-shift maps of the non-visible solar hemisphere measured through a helioseismic technique. Regions with a negative phase shift (a reduction in the travel time of the waves) are potentially associated with the presence of an active region. We use maps obtained from the Joint Science Operation Center (JSOC) repository\footnote{\url{http://jsoc.stanford.edu/ajax/lookdata.html}}, computed using helioseismic holography from Doppler data obtained by the Helioseismic and Magnetic Imager \citep[HMI,][]{Schou+etal2012}. These seismic maps are published in Carrington coordinates with a periodicity of 12 hours. Two kinds of phase-shift maps are made public on the JSOC repository, one computed using a temporal window of 24 hours of Doppler data and another computed using 5 days of the same data.

\subsection{Standard seismic method}\label{SS}

The presence of strong far-side active regions is routinely detected using the Stanford's Strong-Active-Region Discriminator (SARD)\footnote{Further insight on the system can be found at \\
\url{http://jsoc.stanford.edu/data/farside/explanation.pdf}.} on seismic maps processed with five days of HMI Doppler data.
\begin{figure*}[ht]
    \centering
  \includegraphics[width=\textwidth]{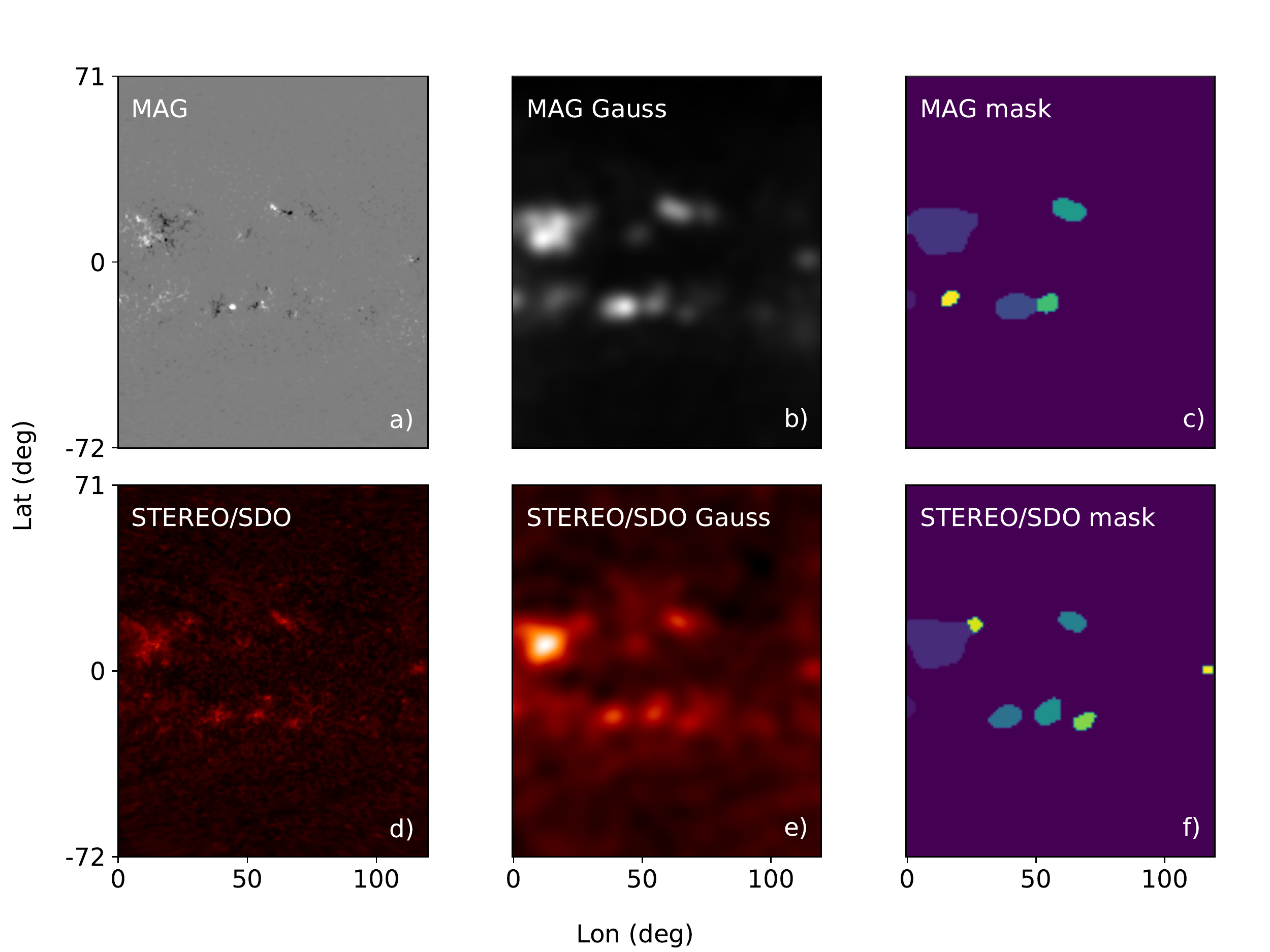}
  \caption{Processing steps performed to calculate the best threshold to compute EUV activity masks. Data from September 22, 2013. a): HMI magnetogram; b): Magnetogram after resizing and applying Gaussian smoothing over the absolute values; c): Final magnetogram mask, using 25 G as threshold for features determination; d): Square-root of STEREO/SDO composite image from the same solar region; e): Region of STEREO/SDO image after applying the square-root transformation, resizing and applying Gaussian smoothing; f): Section of final EUV mask, using the best threshold determined by the date-threshold fit.}
  \label{proc}
\end{figure*}
The method searches for regions where the negative phase-shift is greater than 0.085 rad and calculates 
their corresponding seismic strength ($S$), which is given by the integrated 
negative phase-shift over the area of the region. The area unit used in this calculation is 
the millionth of hemisphere ($\mu$ Hem), so that the seismic strength is given in $\mu$Hem\,rad. A signal is classified as a far-side seismic region if the seismic strength $S$ exceeds a threshold of 400 \citep{Liewer+etal2017}. One of the goals of our study is the evaluation of the performance of the standard seismic method under changes in the selection of the $S$ threshold. We have applied the SARD routine on 5-day phase-shift maps on the analysed range of dates. The routine returns output maps with the same size (in longitude and latitude) as the input seismic map, where each region with $S$ stronger than a user specified threshold is labelled with an integer number. We have extracted far-side active regions maps as a function of the threshold, with $S$ thresholds increasing from 0 to 1275 (including the standard value of 400).

\begin{figure}[ht]
    \centering
  \includegraphics[width=8.5cm]{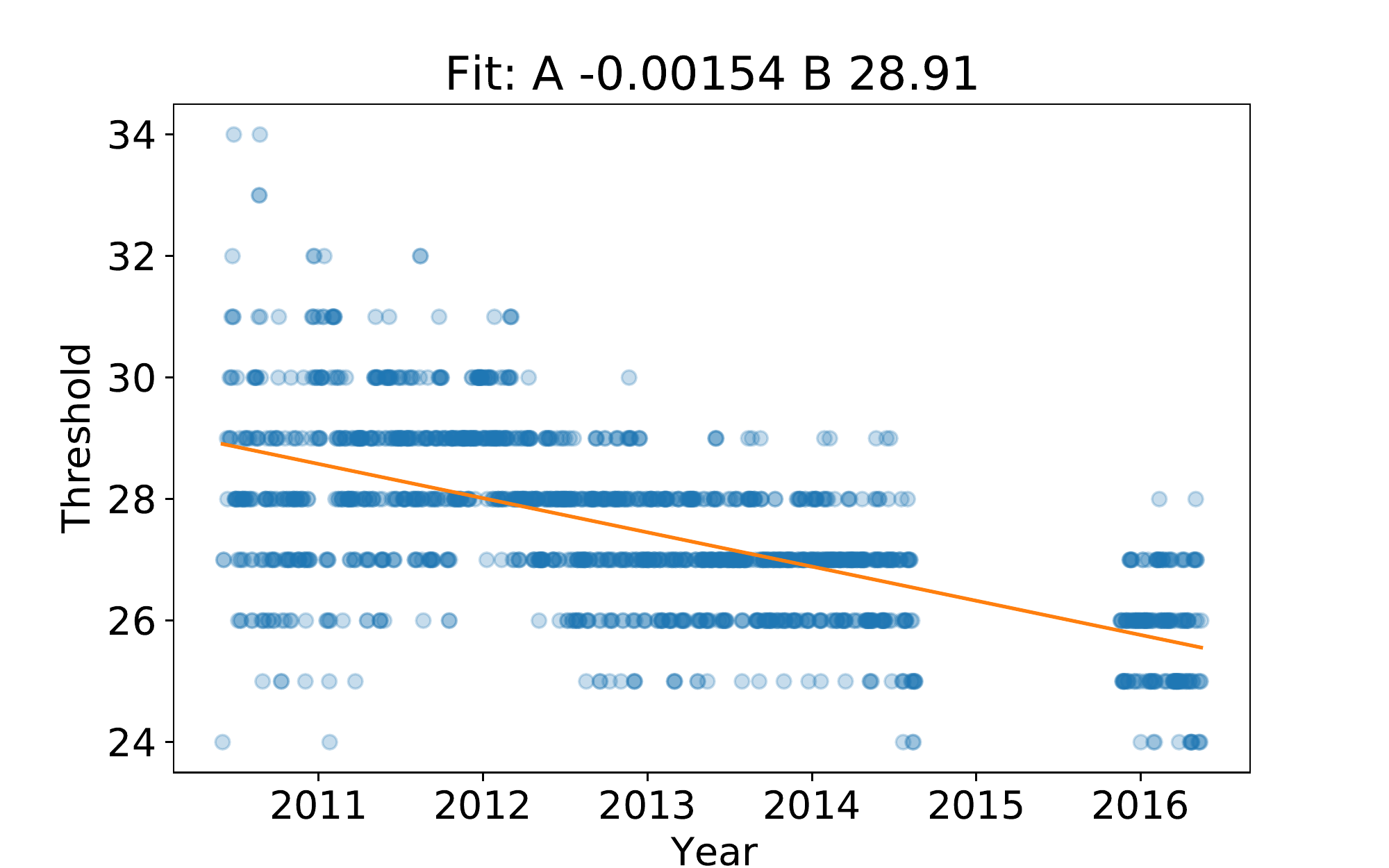}
  \caption{Ideal threshold of the EUV images derived from the comparison of near-side EUV and activity masks as a function of time. Blue dots represent the ideal threshold determined for individual cases. Darker blue regions illustrate a higher concentration of cases. Orange line shows the linear fit of the thresholds. It gives the actual threshold applied to far-side EUV data to compute EUV masks. Fit coefficients can be found on top of the figure. Dates range from June 2010 to May 2016, with a gap of STEREO/SDO composite images from August 2014 to November 2015.}
  \label{fit}
\end{figure}

\subsection{FarNet network}\label{neuralnet}
FarNet\footnote{\url{https://github.com/aasensio/farside}} is a fully convolutional neural network
that follows the standard encoder-decoder U-net architecture \citep{Ronneberger+etal2015}. 
Both the encoder and the decoder are built by the successive applications of convolutional layers, 
batch normalization \citep[BN,][]{Ioffe+etal2015} and rectified linear unit activation functions \citep[ReLU,][]{Nair+Hinton2010}.
The encoder decreases the size of the input images in several steps (while increasing the number
of channels of the tensors) by using max-pooling \citep{Goodfellow-et-al-2016}. The decoder recovers 
the original size by transpose convolutions \citep{Zeiler+Dilip+etal2010}. As standard in the U-net 
architecture, skip connections are used to accelerate training by avoiding vanishing gradients. 
\begin{figure*}[ht]
    \centering
  \includegraphics[width=\textwidth]{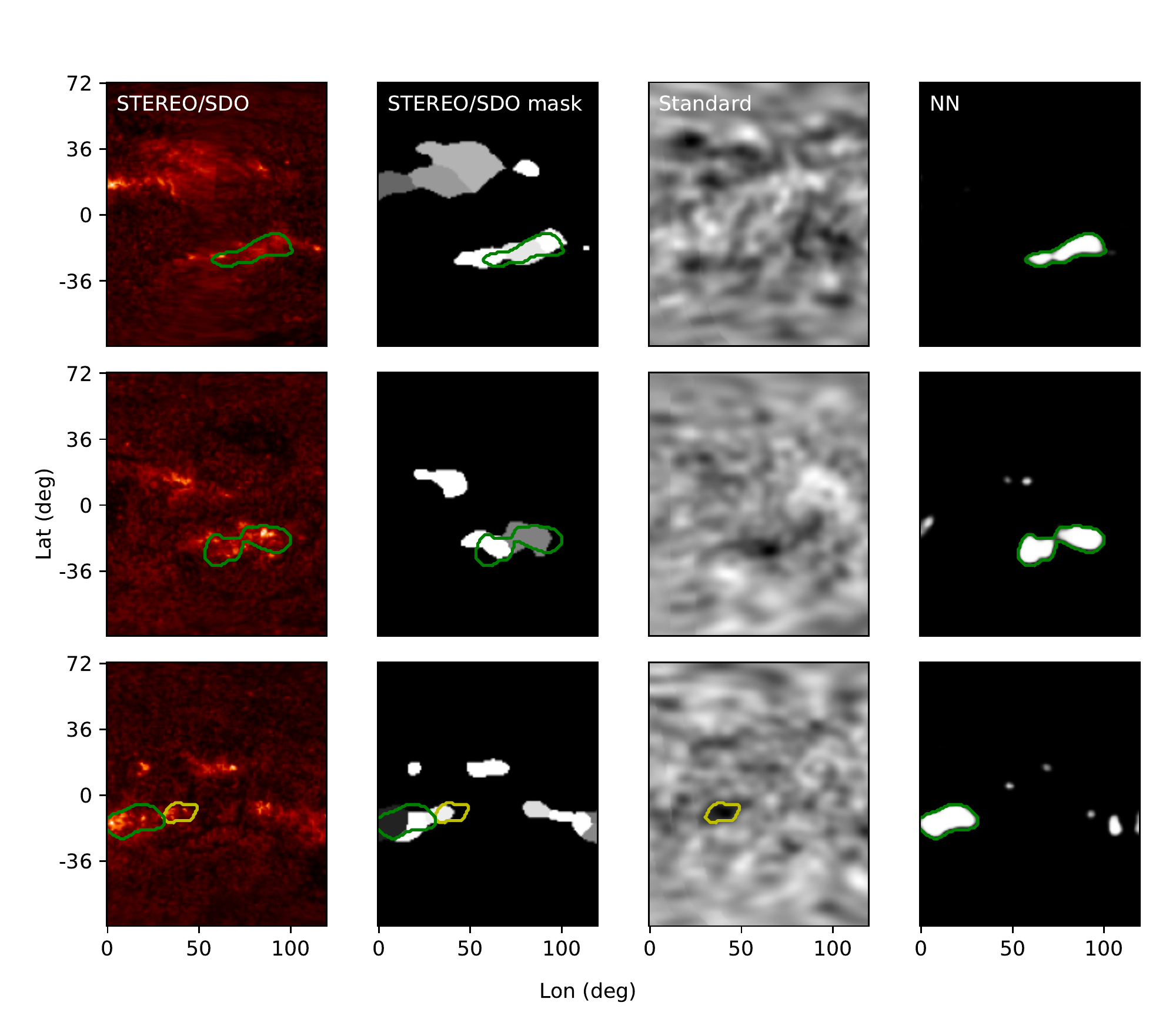}
  \caption{Visual comparison between methods applied on data from 13 April 2011 (top), 8 June 2013 (middle), and 1 April 2014 (bottom). From left to right, square-root of resized STEREO/SDO image, EUV mask, seismic map computed with 5 days of Doppler data, and FarNet output. Yellow and green lines show the standard seismic method detections with $S>400$ and FarNet detections with $P_{\rm i}>100$, respectively.}
  \label{comp} 
\end{figure*}
FarNet takes as input 11 consecutive phase-shift maps computed from HMI Doppler data using helioseismic holography. Each map is obtained from observations acquired during 24 hours and the temporal cadence between consecutive maps is 12 hours. Only a region of the seismic maps, centred on the central meridian of the far-side and ranging 120$^\circ$ in longitude and 144$^\circ$ in latitude, is given to the network as input. The output of the neural network consists of a probability map with the same size as the input maps. To produce such probability maps, the output of the network is equipped with a sigmoid activation function. The network was developed and trained using PyTorch \citep{Paszke+etal2019}.

As explained in \citet{Felipe+Asensio2019}, FarNet was trained to produce binary maps 
which were built by binary segmentation of the near-side HMI magnetograms with a predefined
threshold. The magnetograms were obtained 13.5 days after the central date of the inputs to
allow for half rotation of the solar surface. Active regions that emerged on the near-side were 
carefully removed to keep only those emerged on the far-side that rotated into the visible hemisphere.

Detections in the network output are evaluated by computing the ensuing integrated probability ($P_{\rm i}$). To do so, we search for regions of contiguous pixels with a probability larger than 0.2 and integrate over the area obtained
after applying a Gaussian smoothing of a full width at half maximum of 1.5 pixels to the probability
map. Given that the output map has pixels with an area of 1 deg$^2$, and that probability is dimensionless, the integrated probability has units of deg$^2$. To discriminate true active regions from
potential artefacts, \cite{Felipe+Asensio2019} selected regions with an integrated
probability above $P_{\rm i}$=100. Figure \ref{eleven} shows an example of seismic maps used 
as input for the network and the resulting output, and the corresponding region of the five days cumulative phase-shift map that would be used to compute the standard method prediction for the same date.
\begin{figure*}[ht]
    \centering
  \includegraphics[width=\textwidth]{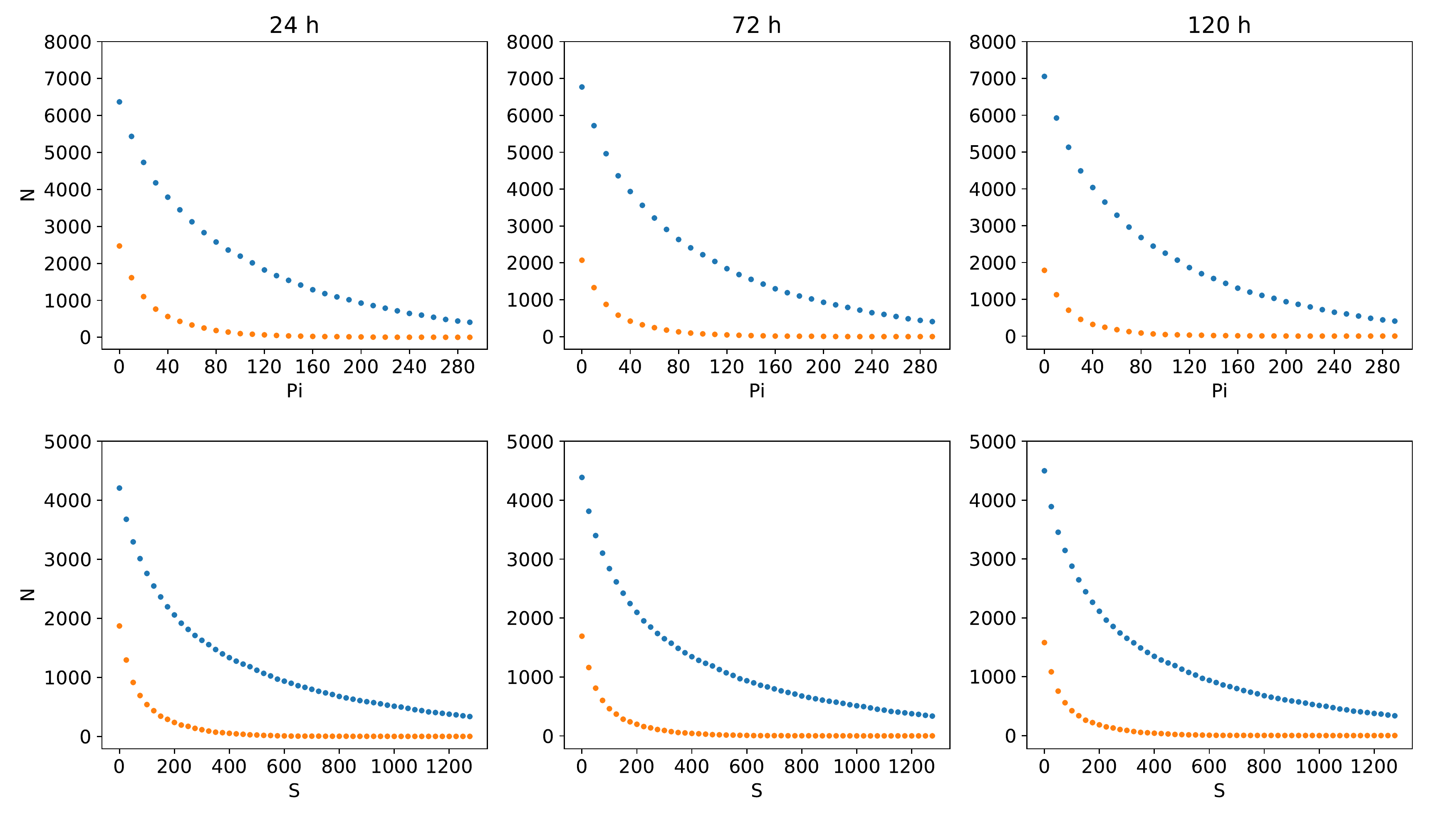}
  \caption{True detections (blue dots) and false positives (orange dots) from FarNet (first row) and from the standard seismic method (second row) over the whole range of dates at study. Each dot corresponds to the detections with $P_{\rm i}$ or $S$ over the X axis value. Each column represents the results of the comparison with 24 h (first column), 72 h (second column), and 120 h (third column) of STEREO data.}
  \label{PiSi}
\end{figure*}
\section{EUV data for far-side activity calibration}\label{STEREO}

As a proxy for the location of active regions, we used 304 {\AA} Carrington maps from the Solar Museum Server of NASA
\footnote{\url{https://solarmuse.jpl.nasa.gov/data/euvisdo_maps_carrington_12hr/304fits/}} \citep{Liewer+etal2014}, 
composed of observations from the Extreme Ultraviolet Imager \citep[EUVI,][]{Wulser+etal2004} onboard STEREO and the instrument Atmospheric Imaging Assembly \citep[AIA,][]{Lemen+etal2012} onboard Solar Dynamics Observatory \citep[SDO,][]{Pesnell+etal2012}. STEREO 304 {\AA} data have been commonly used as an indication of far-side magnetic activity by several works analysing seismic detections of far-side active regions \citep{Liewer+etal2012, Liewer+etal2014, Zhao+etal2019}. Magnetized regions exhibit an increased brightness in the EUV and images in the 304 {\AA} band show a good correlation with magnetic flux maps \citep{UgarteUrra+etal2015}. \cite{Liewer+etal2012} used visual comparison between helioseismic predictions of far-side active regions and STEREO data from February to July 2011 and \cite{Liewer+etal2014} extended the range of study by adding the period from January to April 2012. Later, \cite{Zhao+etal2019} compared 3 months of far-side helioseismic images with STEREO data. Here, we have performed a more statistically significant study by extending the analysis to a larger amount of cases. We use data from the whole temporal period when there was a good STEREO far-side coverage, spanning from April 2011 to May 2016. The far-side coverage by STEREO spacecrafts after late 2014 is not complete, but it is sufficient for our goals since our analysis is focused on the central region of the far hemisphere.   

\subsection{Identification of far-side active regions in EUV data}\label{id}

The comparison of the large set of outputs from the two far-side methods considered in this
work and the STEREO 304 {\AA} images has been automatized. To this end, we transformed the STEREO/SDO data on activity masks (EUV masks from now on). The images were segmented
by considering an intensity threshold, so that every pixel above this threshold is considered to be part of an active region. The ideal EUV thresholds have been determined from the comparison of EUV masks with different thresholds (restricted to the near-side of the composite Carrington maps) with HMI magnetograms acquired simultaneously. The whole procedure is illustrated for one representative case in Fig. \ref{proc}. For the HMI magnetograms, we employed the JSOC series \texttt{hmi.Mldailysynframe\_720s}. The synoptic magnetograms from this series are presented in longitude and the
sine of the latitude, and the first 120$^{\circ}$ in longitude are substituted by a nearly instantaneous magnetogram centred on the central meridian of the visible hemisphere. A total of 1600 STEREO/SDO images and HMI magnetograms were used for the threshold calibration (only the region acquired with SDO/AIA from the STEREO/SDO Carrington maps is actually employed), using one image per day from 2010 June 1 to 2016 May 15. There is a gap of missing STEREO/SDO composite maps from August 2014 to November 2015. 

\begin{figure}[h]
    \centering
  \includegraphics[width=8cm]{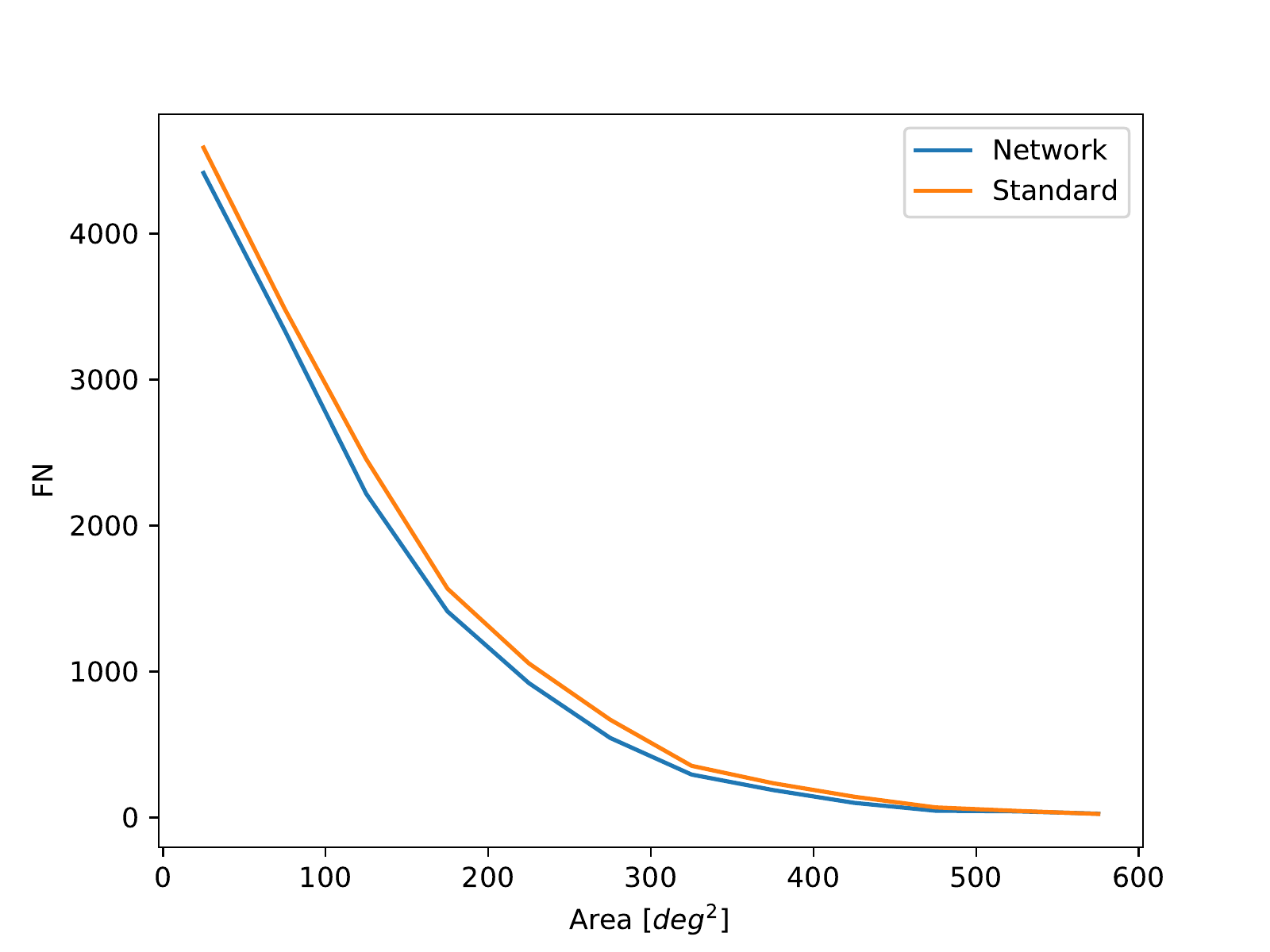}
  \caption{False negatives from the network (blue line) and from the standard method (orange line) as a function of the area of the regions on the activity masks, for the whole range of dates at study. Only the output of the same date than the EUV mask was used for this result, for each method and each EUV mask.}
  \label{FNs}
\end{figure}

In order to properly calibrate the EUV threshold, we first defined activity masks from the magnetograms. For reproducibility purposes, we explain how we did it in the following. Magnetograms were resized to a resolution of 1 deg on both latitude and longitude axis. A Gaussian filter with a standard deviation of three degrees was applied to the absolute value of
the magnetogram signal, to remove small scale magnetic field variations that could difficult the image segmentation (middle panel from top row in Fig. \ref{proc}). From the results of the Gaussian smoothing, activity masks were computed with the IDL routine \texttt{rankdown.pro}, part of the feature tracking software YAFTA\footnote{YAFTA routines found at \url{http://solarmuri.ssl.berkeley.edu//~welsch/public/software/YAFTA}.}. A threshold of 25 G was selected to compute these masks. This value was based on a visual inspection of continuum images of multiple magnetic features, such as sunspots and pores, with the support of the JHelioviewer software \citep{Muller+etal2017}. A example of the resulting magnetogram activity masks is shown in the top-right panel from Fig. \ref{proc}.
\begin{figure*}[ht]
    \centering
  \includegraphics[width=\textwidth]{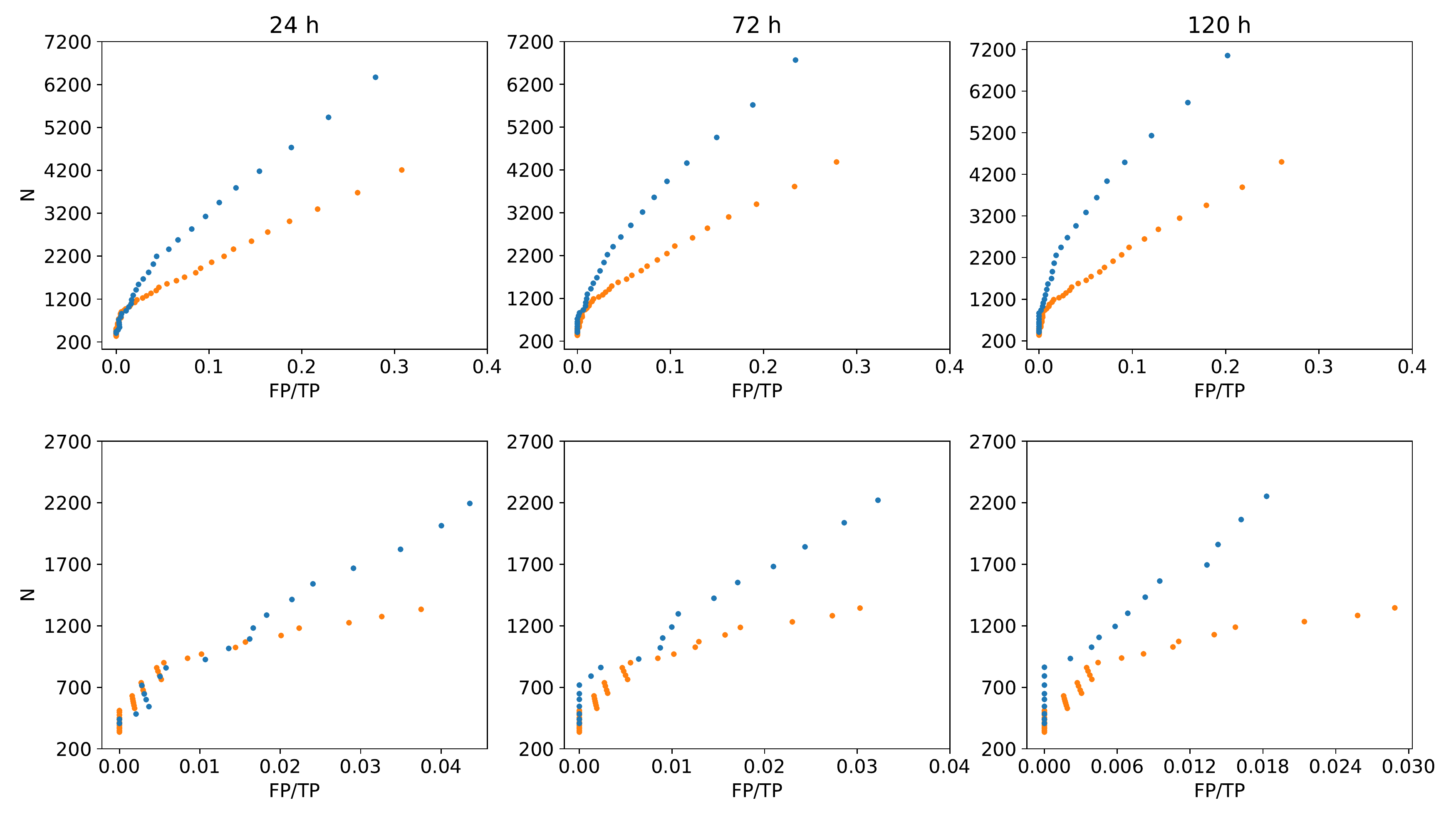}
  \caption{Number of true detections from both methods as a function of the ratio of false positives to total positives. Second row shows a close-up look at the results from thresholds higher than $S=400$ (standard seismic method) and $P_{\rm i}=100$ (FarNet). Each column represents the results of the comparison with 24 h (first column), 72 h (second column), and 120 h (third column) of STEREO data.}
  \label{FP_TP}
\end{figure*}
The EUV data are also processed for the intercomparison, with the consecutive
application of: i) a square-root nonlinear transformation, ii) a downsizing to a resolution of 1 deg in latitude and longitude, iii) a Gaussian filter with standard deviation of three degrees. We then generate 21 EUV masks with the YAFTA routine, with thresholds ranging from 20 to 40 in units of square-root of EUVI DN/s (digital numbers per second). 

Various EUV masks were computed using different thresholds for a sample of dates covering the temporal span between June 2010 and May 2016. The best threshold for each image of the selected EUV sample was chosen such that the ensuing mask was as close as possible to that from the associated magnetogram. This similarity was automatically determined by comparing the features in both mask maps. We found that the ideal EUV threshold depends on the date of the observations (Fig. \ref{fit}). The threshold shows a clear trend with date with negative slope. No clear trend though is found with proxies of activity like the number of sunspots. This trend could be due to the degradation of the STEREO detectors over time. Another reason could be that, although photospheric active regions have a huge effect on the EUV irradiance, they do not explain the whole emission. Approximately 80\% of the total solar irradiance is due to photospheric magnetic activity, but chromospheric and coronal
processes need to be taken into account, being those especially relevant on the EUV part of the spectrum \citep{Domingo+etal2009}. Finally, we compute the EUV masks for the STEREO data set employed for the evaluation of the seismic detections of far-side active regions (April 2011 to May 2016, with a gap in between) using the threshold given by the linear fit shown as an orange line in Fig. \ref{fit}.

\section{Comparison of methods}\label{comparison}

Figure \ref{comp} illustrates the predictions of far-side active regions given by FarNet (last column) and the standard method (second to last column) for three independent cases. They are compared with simultaneous direct observations of the far-side as observed in EUV with STEREO (first column) and the EUV masks constructed for the evaluation of the results (second column).

A visual inspection of Fig. \ref{comp} confirms the well-known fact that the seismic signature of far-side active regions in the visible hemisphere can be employed to detect them. The output from both methods reveals locations where far-side active regions are expected since a reduced travel time is found. Those far-side active regions are confirmed by the enhanced EUV emission measured with STEREO. Interestingly, the outputs from FarNet show a better resemblance with EUV data, including the detection of small regions which are missed by the standard seismic approach.     

We evaluate the performances of both methods as a function of the threshold $P_{\rm i}$ and $S$, using true detections and false positives as performance indicators. 
As explained above, the standard helioseismic method uses data computed from five days of observations (a single seismic map is computed from 5 days of Doppler data) and the neural network approach uses 
data from six days (11 seismic maps computed with observations of 24 hours with a time cadence of 12 hours). In both methods, the use of a longer temporal span improves the signal-to-noise ratio. However, this leads to uncertainties in the determination of the dates where the seismically detected far-side active regions are present. For completeness, we evaluate the reliability of the methods by comparing their prediction with a set of EUV masks. We consider that the time of the prediction is the central date and time of the data employed for the computation (for example, in the case of FarNet it is the time corresponding to 0 h in Fig. \ref{eleven}). This prediction is compared with several EUV masks centered on that date. Three studies were made, comparing the outputs with 24, 72, and 120 hours of EUV data (3, 7, and 11 images, respectively, since the cadence of the employed EUV data is 12 h). A total of 2342 far-side predictions were used in the comparisons. Each of them is considered as an independent case, meaning that they are compared individually with the corresponding EUV masks and that a single active region can be counted several times. In this sense, our statistical approach is similar to that from \cite{Zhao+etal2019}, and differs from \cite{Liewer+etal2017} since we are not tracking active regions.

\subsection{Definitions of true detections and false positives}\label{positives}
A true detection is reported for the output if one of these two conditions is satisfied.
First, the presence of a feature in the EUV masks with a centroid within $\pm15$º in longitude and $\pm5$º in 
latitude of the centroid of any blob on the output \citep[criteria based on the previous work of][]{Liewer+etal2017}. 
Second, if a source on the output of any of the methods has an area larger than 127 deg$^2$, a
superposition between part of the area of the detected regions and the features on the EUV mask will be taken as a true detection. The area selected to take superposition into account is the typical area of a $P_{\rm i}=100$ region on the network outputs. This $P_{\rm i}$ is the threshold chosen for seismic detections in \cite{Felipe+Asensio2019}. If any of the segmented regions of the output satisfies any of the two criteria, it is considered to be a true detection. If none of the criteria are fulfilled, it is considered a false positive. 

\begin{figure*}[ht]
    \centering
  \includegraphics[width=\textwidth]{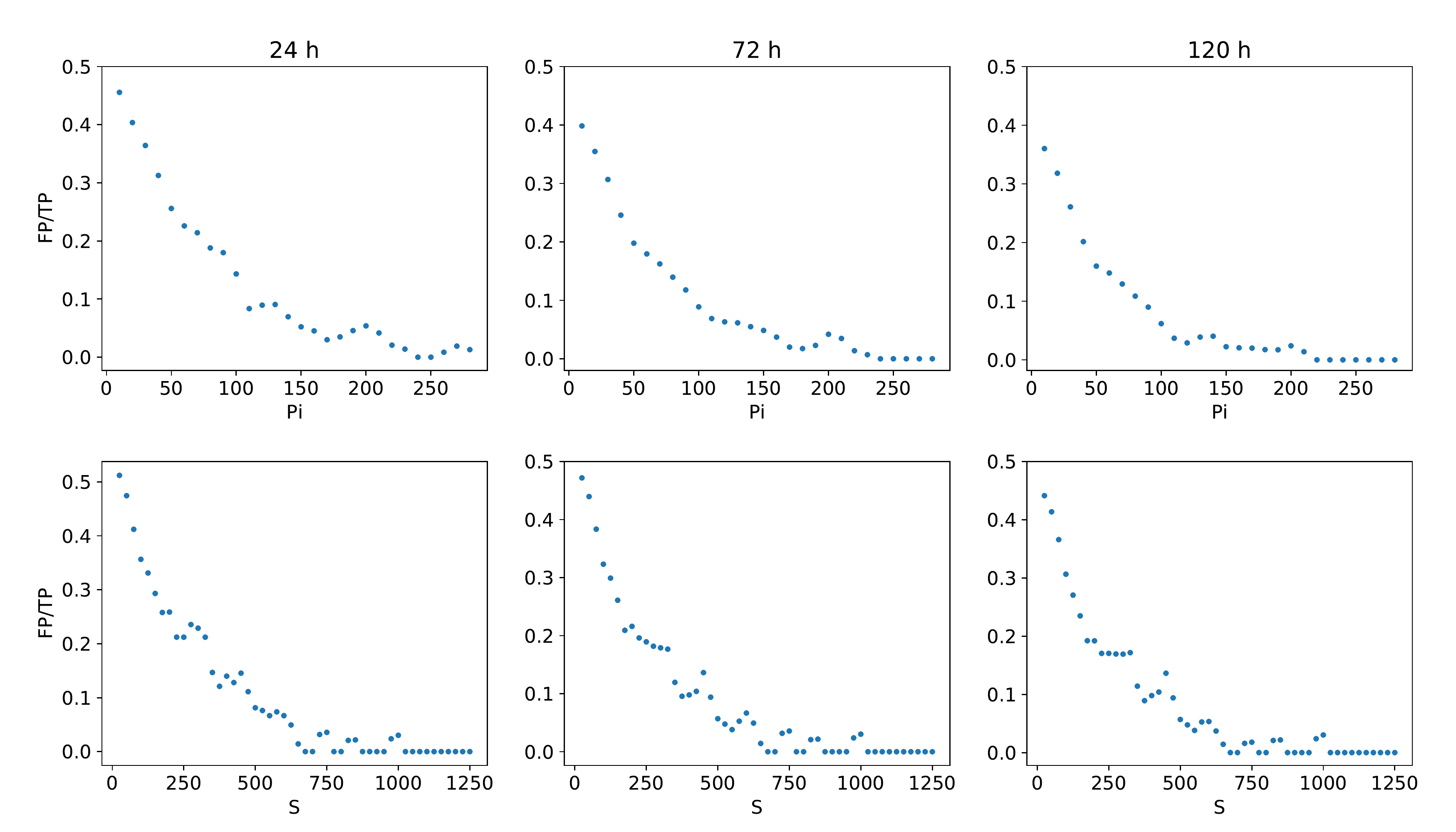}
  \caption{Rate of false positives over all positives as a function of their seismic signature. Each data point represent the ratio of false positives in the ranges [$P_{\rm i}$-10,$P_{\rm i}$+10] (FarNet, top row) and [$S$-25,$S$+25] (standard seismic method, bottom row), with $P_{\rm i}$ and $S$ given by the value in the horizontal axis. Each column represents the comparison results for the studies done with 24 hours (first column), 72 hours (second column), and 120 hours (third column) of STEREO data.}
  \label{cajas}
\end{figure*}

\begingroup
\renewcommand*{\arraystretch}{1.3}
\begin{table}
\caption{Detections, false positives and false positives percentage for the 24 hours range study, with $S>400$ for the seismic method (SS) and for $P_{\rm i}>113$ for the neural network approach (NN).}
\label{table:1}
\centering 
\begin{tabular}{l c c}
\hline
 & Detections & False Positives \\
\hline
   SS $S>400$ & 1334 & 52 (3.75\%)\\
   NN $P_{\rm i}>113$ & 1958 & 76 (3.74\%)\\
\hline
\end{tabular}
\end{table}
\endgroup

\subsection{Results}\label{results}

Figure \ref{comparison} shows the results of applying the standard method and the network to seismic data of three independent dates. For those three days, the network shows a greater sensitivity, and all strong active regions detected by the network ($P_{i}$>100) are associated with EUV emission signatures. Of the three cases illustrated in Fig. \ref{comparison}, only one strong active region (S>400) is detected by the standard method.

Figure \ref{PiSi} shows the number of true detections and false positives for each method. The upper panels
show the results for FarNet whereas the lower panels show those for the standard seismic method.
All plots indicate the number of regions above a certain threshold in $P_{\rm i}$ or $S$, depending
on the method. Results are shown for the comparison with windows of 24, 72, and 120 h of EUV data. These figures clearly show that,
once a sufficiently small false detection rate is reached, the number of true detection for FarNet
greatly exceeds those of the standard seismic method. For instance, we find that $P_{\rm i}=120$ produces
a negligible amount of false positives for FarNet, giving near to 2000 true detections in the 24 h window.
In the same conditions, a negligible false positive rate is reached for $S=400$ for the standard seismic
method, which then produces less than 1500 true detections.

Figure \ref{FP_TP} shows a different view of the results for a better
inter-comparison of the two methods. We display the number of true detections for each method as
a function of the ratio of false positives (FP) to total positives (TP, true detections plus false positives). Each point 
on the figure is associated with a $P_{\rm i}$ or $S$ threshold in particular. The lower panels are close-ups for the cases with $P_{\rm i}>100$ and $S>400$, which are the thresholds previously suggested for each method \citep{Liewer+etal2017,Felipe+Asensio2019}. Lower-left panel from Fig. \ref{FP_TP} shows that for larger values of $P_{\rm i}$ and $S$ both methods return similarly good results. In the rest of the range of false positives ratio (and in the whole range, in the case of windows of 72 and 120 h of EUV data, middle and right panels) it is clear that the FarNet performance is superior than the standard seismic method. This proves that FarNet is able to reliably detect smaller activity signatures than the standard method. 

Table \ref{table:1} shows a comparison of both methods employing the threshold defined in previous studies for the standard method ($S=400$) and that which lead to the closer percentage of false positives for the network results ($P_{\rm i}=113$), for the 24 hours study. For those thresholds, FarNet supposed an improvement of $\sim$ 47\% in the number of true detections. 

Figure \ref{cajas} illustrates the rate of false positives as a function of the signature ($P_{\rm i}$ or $S$) of the detected active regions. In this figure, the horizontal axis represents the centre of the range employed for each individual measurement. In the case of FarNet (top row), each dot indicates the ratio of false positives for the set of detections with $P_{\rm i}$ in the range [$P_{\rm i}$-10, $P_{\rm i}$+10], whereas the bottom panels illustrate the same quantity but for seismically detected active regions in the range [$S$-25, $S$+25]. This representation differs from previous analyses (Figs. \ref{PiSi} and \ref{FP_TP}) since the considered detections for each data point are restricted to those inside the defined bin, instead of including all the active regions with $P_{\rm i}$ or $S$ above the corresponding threshold. The results from Fig. \ref{cajas} provide an indication of the reliability of an independent far-side seismic detection (characterized by its $P_{\rm i}$ and/or $S$) supported by the statistical analysis of similar detections over approximately four years of data. 

Figure \ref{FNs} shows the false negatives of both methods as a function of the area of the features on the EUV masks,  for a comparison of each EUV mask with the output related to the same date. False negatives are defined as regions where EUV emission is found, but with no association to a seismically detected active region. Detections were claimed for regions on the outputs with $P_{\rm i}$>113 or S>400. These false negatives were studied for intervals of 100 deg$^2$ of area, from 0 to 1100. We found that, for regions on the EUV masks with an area over 600 deg$^2$, both the network and the standard method can detect virtually all the sources. This is in agreement with \citet{Zhao+etal2019}, who found a similar area threshold for the complete detection of EUV active regions. For smaller regions, the network leaves less regions without detection, as expected from previous results.

\section{Discussion and conclusions}\label{disc}

Since the pioneering works of \cite{Lindsey+Braun2000} and \cite{Braun+Lindsey2001}, the detection 
of far-side active regions based on the analysis of the oscillations in the visible hemisphere 
has been a common approach. These predictions are regularly published as one of the data 
products from SDO and the Global Oscillation Network Group \citep[GONG,][]{Harvey+etal1996}. 
Later works have subsequently validated the method and evaluated its performance. Due to 
the absence of direct far-side observations during the first years of this century, those 
first analyses compared the seismic predictions with observations taken after the probed solar 
regions had rotated into the visible solar hemisphere. These studies revealed the relation between 
the seismic signature and the magnetic field strength \citep{GonzalezHernandez+etal2007} and found 
that 40\% of the total active regions appearing at the east limb of the Sun can be detected with 
a confidence level higher than 60\% \citep{GonzalezHernandez+etal2010}. With the advent of 
the STEREO spacecrafts, direct observations of the far-side have been available, allowing a better
evaluation of the reliability of the seismic predictions. These methods have shown a remarkable
performance to detect strong far-side active regions 
\citep{Liewer+etal2014, Liewer+etal2017, Zhao+etal2019}.

In this study, we have evaluated two different approaches to identify active regions in 
holography far-side seismic maps: the standard method of computing the seismic strength of the 
regions and the recently developed neural network FarNet. The predictions from those two approaches 
have been compared with direct EUV far-side observations acquired with STEREO. In both cases, we 
have quantified the results as a function of the threshold selected for the identification of active
regions. Our study benefits from the analysis of a larger sample than previous works. The evaluation 
of the standard method confirms the reliability of this approach to detect strong active regions. 
Our results support the selection of $S=400$ as threshold for the identification of far-side active
regions since it provides a high number of detections with a low risk of false positives 
(around 4\%, see Table \ref{table:1}).

\cite{Felipe+Asensio2019} validated the neural network FarNet using a limited sample 
of STEREO data. That sample covered a time period during the solar minimum (and, thus, 
with low availability of active regions) and without full STEREO coverage of the far-side.
\cite{Felipe+Asensio2019} showed the promising results provided by FarNet, and here we 
have confirmed the potential of the method using a greatly improved statistical sample. A 
comparison of FarNet results with the standard approach shows that both methods are similarly 
good for strong active regions ($S\gtrsim500$). In the case of weaker active regions, FatNet 
exhibits a significantly better performance than the standard method. For thresholds (in $S$ 
and $P_{\rm i}$) giving a similar success rate, FarNet manages to identify a much higher 
number of active regions (Fig. \ref{FP_TP}). 

These results show the potential of FarNet to contribute to a great number of applications. 
The detection of far-side active regions is fundamental for space weather studies, including 
the forecast of the spectral irradiance \citep{Fontenla+etal2009} and the solar 
wind \citep{Arge+etal2013}. One way to probe far-side active regions is through EUV far-side 
imaging. STEREO spacecrafts have lead to massive amounts of EUV data from the far-side 
hemisphere since their launch in 2006. Using this data, \cite{Kim+etal2019} were able 
to infer far-side magnetograms with fair success with a machine learning tool. This technique 
can be very helpful while EUV far-side data are available, but unfortunately far-side 
observations are a limited resource. STEREO spacecrafts are returning to the Earth-side 
and STEREO-B has stopped transmitting. ESA mission Solar Orbiter 
\citep[launched on February 2020,][]{Muller+etal2013} will provide far-side 
magnetograms, but only during some periods of its orbit. Seismic inference will remain 
a key tool over the next decades since it will be the only approach to obtain a 
constant monitoring of the solar far-side hemisphere.

In this context, there is a growing interest in improving the methods for the seismic 
detection of far-side active regions. Recent developments have taken steps in this 
direction by exploring various time-distance measurement schemes with more multiskip waves 
\citep{Zhao+etal2019} and by employing machine learning tools \citep{Felipe+Asensio2019}. 
Here, we have further analysed the network presented in the latter work, proving its promising 
results. This study is restricted to active regions around the centre of the far side (up to 60$^{\circ}$ from the far-side centre in longitude) and our statistics do not discern the locations of the regions. However, one would expect a better performance near the centre of the far side than near the limbs. Future efforts will be devoted to evaluating the sensitivity of both methods with the closeness of the active regions to the limb. Additionally, there is still room for the improvement of FarNet. The training of the network was made using,
as outputs, binary magnetograms of the near-side taken 13.5 days after the prediction date. That 
is, they are not co-temporal to the seismic images. This limitation can potentially be bypassed 
by employing the STEREO EUV masks defined in this work and/or future far-side magnetograms acquired 
with Solar Orbiter for the training. We are actively working on this line and the
preliminary results look promising. Other possible developments include the direct use of 
Doppler maps as inputs, instead of phase-shift seismic maps, and the design of neural 
networks capable of returning an estimation of far-side magnetic flux and/or magnetograms.

\begin{acknowledgements}
We thank P. C. Liewer and collaborators for making publicly available the composite STEREO/EUVI and SDO/AIA maps necessary to carry out this research. We thank C. Lindsey and D. C. Braun for providing the Stanford’s Strong-Active-Region Discriminator routine to compute predictions of active regions on the far-side. This work was supported by the State Research Agency (AEI) of the Spanish Ministry of Science, Innovation and Universities (MCIU) and the European Regional Development Fund (FEDER) under grant with reference PGC2018-097611-A-I00 and by Consejería de Economía, Conocimiento y Empleo del Gobierno de Canarias and the European Regional Development Fund (ERDF) under grant with reference PROID2020010059.
We acknowledge the community effort devoted to the development of the following 
open-source packages that were
used in this work: \texttt{numpy} \citep[\texttt{numpy.org},][]{numpy20}, 
\texttt{matplotlib} \citep[\texttt{matplotlib.org},][]{matplotlib}, \texttt{PyTorch} 
\citep[\texttt{pytorch.org},][]{Paszke+etal2019}, and \texttt{SunPy} 
\citep[\texttt{sunpy.org},][]{sunpy_community2020}.
\end{acknowledgements}

 \bibliographystyle{aa} 
 \bibliography{biblio.bib}

\end{document}